\documentclass[aps,prx,twocolumn,floatfix,longbibliography,nofootinbib,superscriptaddress]{revtex4-1}
\usepackage[utf8]{inputenc}
\usepackage{natbib}
\usepackage{graphicx}
\usepackage{xcolor}
\usepackage[tbtags]{amsmath}
\usepackage[colorlinks, linkcolor=blue]{hyperref}
\hypersetup{colorlinks,allcolors=black}
\usepackage{amssymb}
\usepackage{gensymb}
\usepackage{float}
\usepackage{amsmath}
\usepackage{tabularx,graphicx}
\usepackage{epstopdf}
\usepackage{latexsym}
\usepackage{color, colortbl}
\usepackage{psfrag}
\usepackage{bbm}
\usepackage{bm}
\usepackage{titlesec}
\usepackage{dsfont}
\usepackage{feynmp}
\usepackage{slashed}
\usepackage{multirow}
\usepackage[normalem]{ulem}

\def \R {{\cal{R}}}

\def \beq {\begin{eqnarray}}
\def \eeq {\end{eqnarray}}
\def \tn {\textnormal}

\begin{document}
\title{Continuous Mott transition in moir\'e semiconductors: \\role of long-wavelength inhomogeneities}
\author{Sunghoon Kim}
\affiliation{Department of Physics, Cornell University, Ithaca, New York 14853, USA.}
\author{T. Senthil}
\affiliation{Department of Physics, Massachusetts Institute of Technology, Cambridge, Massachusetts 02139, USA.}
\author{Debanjan Chowdhury}
%\email{debanjanchowdhury@cornell.edu}
\affiliation{Department of Physics, Cornell University, Ithaca, New York 14853, USA.}

\begin{abstract}
Recent experiments in moir\'{e} transition metal dichalcogenide materials have reported the observation of a continuous bandwidth-tuned transition from a metal to a paramagnetic Mott insulator at a fixed filling of one electron per moir\'{e} unit cell. The electrical transport measurements reveal a number of puzzling features that are seemingly at odds with the theoretical expectations of an interaction induced, but disorder-free, bandwidth-tuned metal-insulator transition. In this work, we include the effects of long-wavelength inhomogeneities, building on the results for a continuous metal-insulator transition at fixed filling in the clean limit. We examine the effects of meso-scale inhomogeneities near the critical point on transport using the framework of random resistor networks, highlighting the salient differences from a simple percolation-based picture. We place our results in the context of recent and ongoing experiments. 
\end{abstract}

\maketitle

{\it Introduction.-} Electrical transport properties of metals and insulators at low-temperature are vastly different. The finite, but small, low-temperature resistance in conventional metals with weak disorder is due to elastic scattering off impurities. On the other hand, electrical transport in insulators with a finite charge-gap is nearly frozen out. Across a continuous metal-insulator transition (MIT), the theoretical mechanism leading to the dramatic suppression of the electrical conductivity is a question of fundamental interest, made especially difficult by the complex interplay of interaction and disorder effects. Recent experimental advances in tunable moir\'e materials have led to the demonstration of a bandwidth-tuned continuous quantum phase transition (QPT) between a metal and an interaction-induced Mott insulator at a fixed (commensurate) filling  \cite{ghiotto_quantum_2021,li_continuous_2021}. 

We focus here on one of these experiments \cite{li_continuous_2021}, which studied a continuous MIT in a transition metal dichalcogenide (TMD) heterobilayer as a function of an external displacement field at half-filling. Approaching the transition from the insulating side, the charge-gap vanishes continuously and is not associated with long-range magnetism (at least down to $5\%$ of the Curie-Weiss scale). Moreover, the spin-susceptibility evolves smoothly across the MIT suggesting that the Mott insulator has an abundance of low-energy spinful excitations, and displays a ``Pomeranchuk" effect as a function of increasing temperature. The temperature-dependent resistivity traces exhibit beautiful scaling collapse into two sets of insulating and metallic-like curves, with the possible exception of the immediate vicinity of the critical point suggesting on the important role played by inhomogeneities. One possible theory \cite{AG04} that can help explain much of this data is associated with the transition from a Fermi-liquid metal to a paramagnetic Mott insulator with a Fermi surface of electrically neutral spinons \cite{senthil_theory_2008}. Across such a QPT, the spin-susceptibility evolves smoothly as the electronic Fermi surface morphs into the spinon Fermi surface, while the quasiparticle residue and charge-gap vanish continuously upon approaching the critical point from either side. A number of complimentary theoretical approaches have also been used to analyze other aspects of a MIT and insulating regimes \cite{SDS20,AM21,SDS21, Millis-HF, Millis-DMFT} in this system, respectively. The role of disorder has been analyzed starting from a different point of view across (but not necessarily, continuous and bandwidth-tuned) MITs \cite{VD09,ghaemi,VD21,SDS22}; see also \cite{RMP1,RMP2,spivak}.

\begin{figure}
\centering
\includegraphics[width=0.49\linewidth]{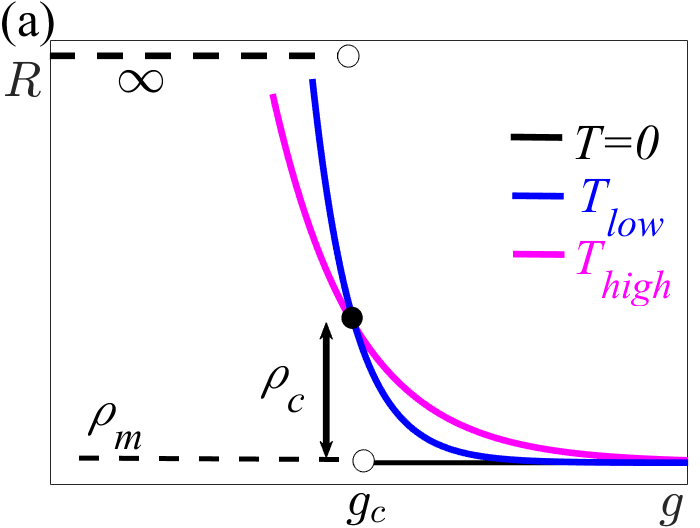}
\includegraphics[width=0.49\linewidth]{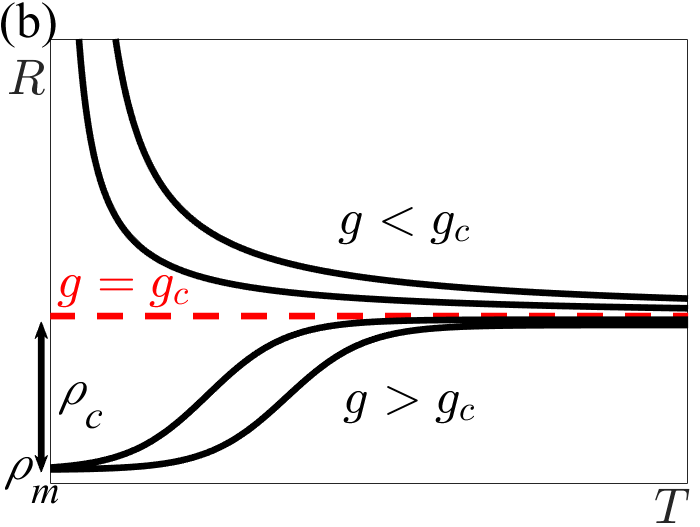}
\includegraphics[width=0.49\linewidth]{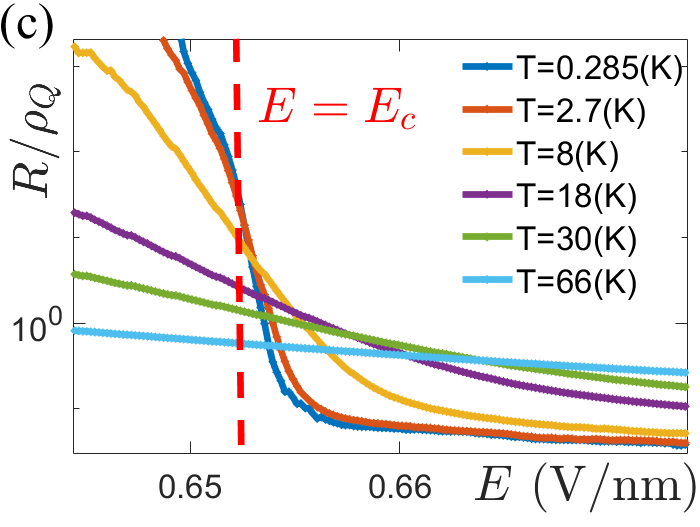}
\includegraphics[width=0.49\linewidth]{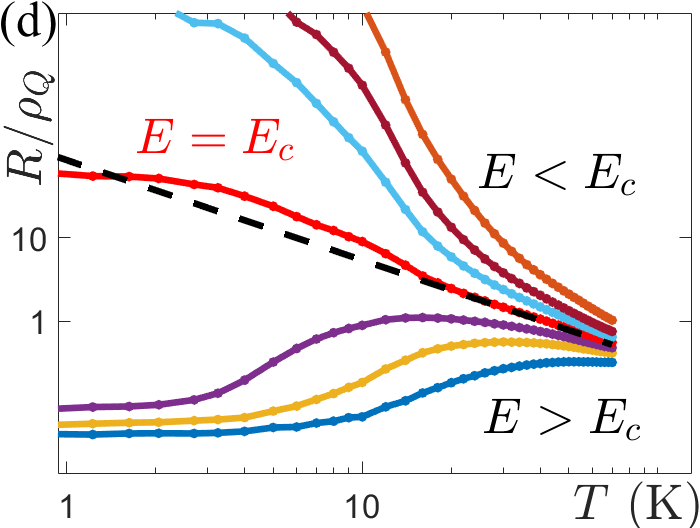}
\caption{(a) and (b) Theoretical expectation for resistivity evolution in the clean limit as a function of temperature ($T$) and tuning parameter ($g$), across a bandwidth tuned MIT at $g=g_c$. There is a universal jump in the resistivity, $\rho_c=\mathcal{R}\rho_Q$, at $g=g_c$. For $T>0$, the resistivity traces cross at a single point. (c) Experimental results for the resistance as a function of out-of-plane electric field, $E$, for $T>0$ exhibit multiple crossing points \cite{li_continuous_2021}. (d) At the critical field $E=E_c$, the resistance exhibits an unusual power-law behavior (black dashed line) $R\sim T^{-\alpha}$ with $\alpha \approx 1.2$.}
\label{fig1}
\end{figure}

Despite the broad agreement between the experiment and theory, there are a number of mysteries associated with the transport measurements, especially when placed in the context of the theory for a continuous metal to paramagnetic Mott insulator transition. In the absence of microscopic inhomogeneities, the theoretical expectation for the resistivity evolution as a function of external tuning parameter and temperature is as shown in Fig.~\ref{fig1}(a) and (b). At asymptotically low temperatures, the resistivity is accompanied by a universal ``jump" of magnitude, $\rho_c = \R \rho_Q~~(\rho_Q=h/e^2)$, across the MIT. $\R$ here is a universal number whose exact numerical value is determined by the strongly coupled field-theory describing the critical point \cite{senthil_theory_2008,witczak-krempa_universal_2012,PhysRevX.12.021067}. Moreover, as a function of increasing temperature, all resistance traces should cross at a {\it single} point \cite{witczak-krempa_universal_2012}. On the other hand, the experiments reveal the following anomalous features: (i) the transport data reveals no sign of a resistivity-jump; instead, the low-temperature resistance grows dramatically upon approaching the putative critical point from the metallic side (Fig.~\ref{fig1}(c)); (ii) the resistivity traces exhibit {\it multiple} crossings at different values of the tuning parameter as a function of increasing temperatures (Fig.~\ref{fig1}(c)); (iii) the normalized sheet-resistance, $R(T)/R_c(T)$, exhibits scaling collapse into two families of mirror symmetric curves, where $R_c(T)$ represents an unusual power-law resistance at the critical point (Fig.~\ref{fig1}(d)). It is natural to ask if resolving these discrepancies between theory and experiment requires an entirely new starting point. 

In this letter, we will demonstrate that all of the unexpected features associated with electrical transport are likely tied to the effects of meso-scale inhomogeneities near the continuous MIT. We do not describe the microscopic origin of such inhomogeneities here, which likely arise due to a variety of reasons including some combination of Coulomb impurities, twist-angle disorder etc. In our model, the sample consists of ``puddles", with a characteristic mean bandwidth defined locally within each puddle.{\footnote{The variations in local twist-angle and the size of the moir\'e unit cell can also induce chemical-potential disorder. However, we will consider the simpler situation with only a local bandwidth disorder, which already helps us address the experimental puzzles.}} Electrical transport is then best described in terms of a random network of such puddles. However, as we explain below, a naive bond-percolation picture in terms of only metallic and insulating links is insufficient to account for the experimental phenomenology. The intrinsic correlation effects that lead to additional temperature-dependent fraction of puddles with ``critical" resistance ($\rho_c\sim O(h/e^2)$) lead to a modified percolative transition. 

{\it Model.-} Let us first consider a simplified picture at $T=0$. We consider an external parameter $g (\sim \tn{Bandwidth/Interaction})$ driving the QPT across $g=g_c$. As a result of the long-wavelength disorder, the system is effectively described by a network of resistors corresponding to either metallic (\(g>g_c\)) or insulating (\(g<g_c\)) puddles. For a given {\it mean} value of the tuning parameter, $g_0$, we assume that the distribution of resistors is given by, $P_{\sigma}(g,g_0)=e^{-(g-g_0)^2/{\sigma}^2}/\sqrt{\pi\sigma^2}$, where $\sigma$ represents a characteristic disorder strength. Henceforth, \(g\) will denote a unitless parameter (i.e. \(g/\sigma \rightarrow g\)). At $T=0$, the metallic resistors are assumed to have a small residual resistance, $\rho_m\ll \rho_Q$, due to weak elastic impurity scattering while the insulating resistors have infinite resistance (Fig.~\ref{fig1}(a)). The system is then effectively described by bond-percolation and the conductance of this random resistor network (RRN) can be computed using the transfer-matrix method \cite{derrida_transfer_1982,han_computer_1998}.

\begin{figure}
\centering
\includegraphics[width=0.485\linewidth, height=0.4\linewidth]{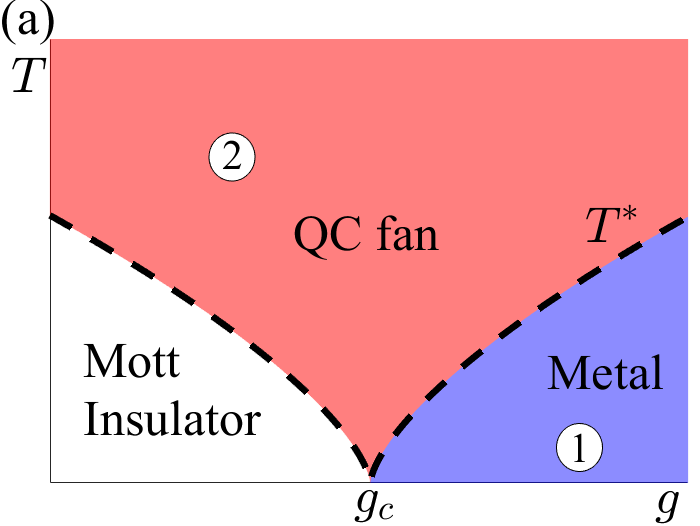}
\includegraphics[width=0.5\linewidth,height=0.395\linewidth]{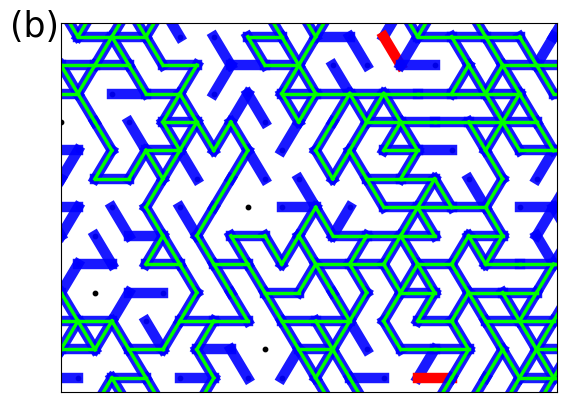}
\includegraphics[width=0.495\linewidth, height=0.4\linewidth]{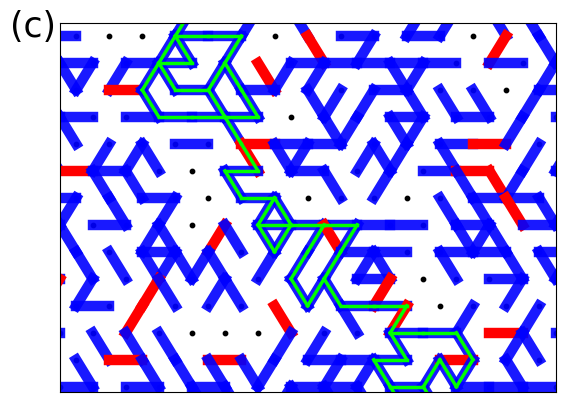}
\includegraphics[width=0.49\linewidth, height=0.42\linewidth]{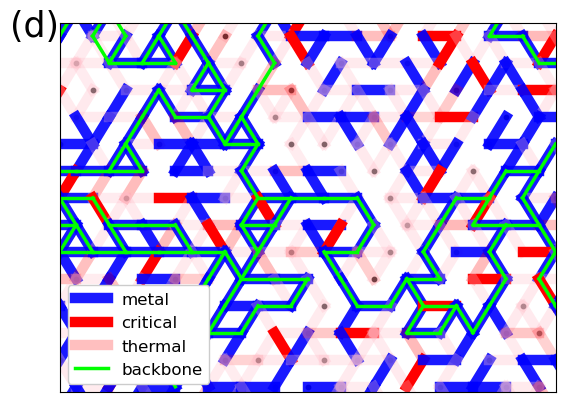}

\caption{(a) A simplified phase diagram for a continuous MIT showing a metallic, (Mott) insulating, and critical regime, where $T^* \sim |g-g_c|^{\nu z}$ denotes a crossover scale with $\nu,~z$ the critical exponents for the clean theory \cite{senthil_theory_2008}. (b) A typical snapshot of the percolating current path (green) consisting of metallic resistors (blue) for a point labelled as $\textcircled{1}$, deep in the metallic regime. (c) The percolating backbone for the point $\textcircled{2}$ contains metallic and critical resistors (red). (d) For the same point $\textcircled{2}$, introducing a finite resistance due to the thermally activated resistors (light-red) contributes to the percolating backbone.}
\label{fig2}
\end{figure}

Next we turn to the role of a finite temperature, which is incorporated  within our model via two distinct mechanisms. For the insulating puddles, we assume that their resistance has an activated form, $\rho = \rho_c e^{\Delta_g/2k_BT}\gtrsim \rho_Q$, where $\Delta_g=\Delta_0|g-g_c|^{\nu z}$ is the ``local" gap-size within the puddle, with a characteristic energy \(\Delta_0\). For our numerical computations and comparison with experiments, we will assume that the puddle itself is locally described by the clean theory associated with a QPT from a metal to a paramagnetic Mott insulator with spinon Fermi surface; the associated correlation length exponent, $\nu\sim0.67$, and dynamical exponent, $z=1$ \cite{senthil_theory_2008}. Additionally, as a function of increasing temperature, the system will crossover into the quantum-critical (QC) regime, which is controlled by the fixed-point associated with the clean theory (Fig.~\ref{fig2}a). %($T\gtrsim T^{*}\sim |g-g_c|^{\nu z}$) 
As noted earlier, transport in this regime is dominated by critical resistors, $\rho_c = \R \rho_Q$, and the volume fraction of puddles with a critical resistance, $f_{\tn{critical}}$, is controlled by the width of the QC fan at a given temperature (Fig.~\ref{fig2}(a)). %As we will demonstrate below, 
Once we include both temperature-dependent effects, the numerical calculations will lead to a remarkable agreement with the experimental phenomenology.

To understand how the metallic, insulating and critical puddles contribute to the current carrying paths, let us consider two representative points, $(g_0 ,T)$, on a simplified phase diagram (Fig.~\ref{fig2}(a)). We show a snapshot of a typical percolating current path (green network) in Fig.~\ref{fig2}(b) for the point labelled $\textcircled{1}$, at low temperature and deep in the metallic regime, which involves almost entirely metallic resistors (blue) and practically no critical resistors (red). In this low temperature limit, the total fraction of resistors, $(f_{\tn{critical}}+f_{\tn{metal}})\approx f_{\tn{metal}}\gg p_c$, where $p_c$ is the percolation threshold on the triangular lattice. We note that the (bond) percolation threshold is defined as the fraction of connected bonds in a network, at which an infinitely-large cluster forms and the percolation sets in \cite{stauffer2018introduction}. Thus, deep in this metallic regime, the critical resistors do not play any significant role. We now turn to the point labelled $\textcircled{2}$, a point at high temperature in the QC regime, where we assume that $(f_{\tn{critical}}+f_{\tn{metal}})\rightarrow p_c^+$ approaches the percolation threshold from above. Let us first ignore thermal activation for the insulating links. With increasing temperature, the width of the QC fan becomes larger thereby increasing $f_{\tn{critical}}$. Thus, the current carrying ``backbone" now inevitably contains critical resistors (Fig.~\ref{fig2}(c)). As long as these critical resistors are located at singly-connected bonds along the backbone, it can lead to a significant increase in the resistance by $\sim O(\rho_Q)$. If we now introduce a thermally activated contribution to the resistance due to the insulators, the current path will also include those insulating resistors with a small gap (light-red) in the vicinity of the critical point (Fig.~\ref{fig2}(d)). Clearly, as one approaches the percolation threshold, the overall conductivity will be determined by the non-trivial connectivity along the backbone, which we address next using a more quantitative analysis. 

\begin{widetext}

\begin{figure}[h]
\centering
\includegraphics[width=0.32\linewidth]{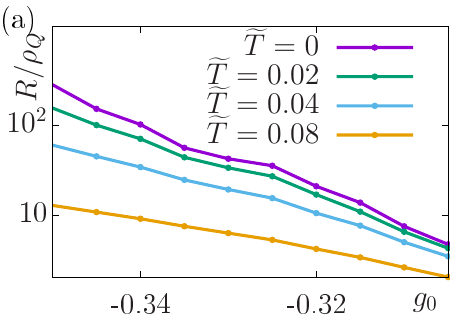}
\includegraphics[width=0.32\linewidth]{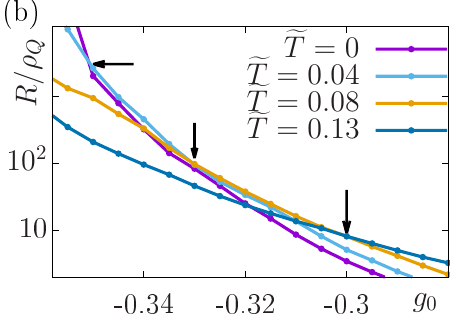}
\includegraphics[width=0.32\linewidth]{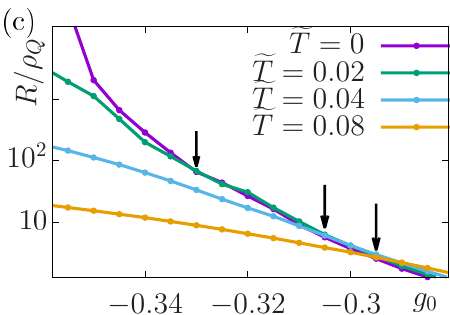}
\includegraphics[width=0.32\linewidth]{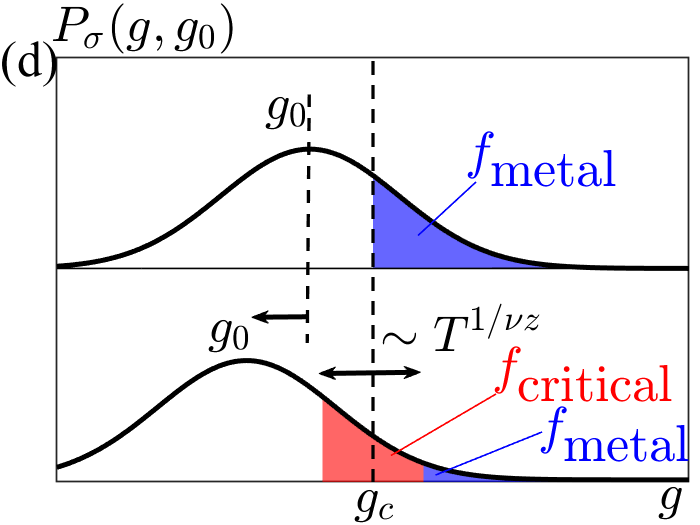}
\includegraphics[width=0.32\linewidth, height=0.24\linewidth]{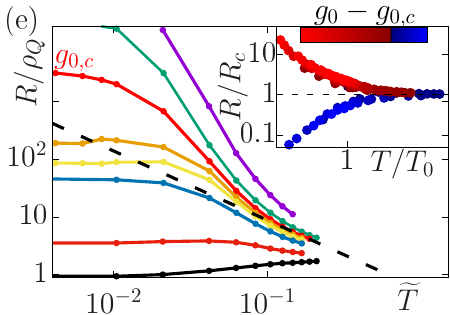}
\includegraphics[width=0.32\linewidth]{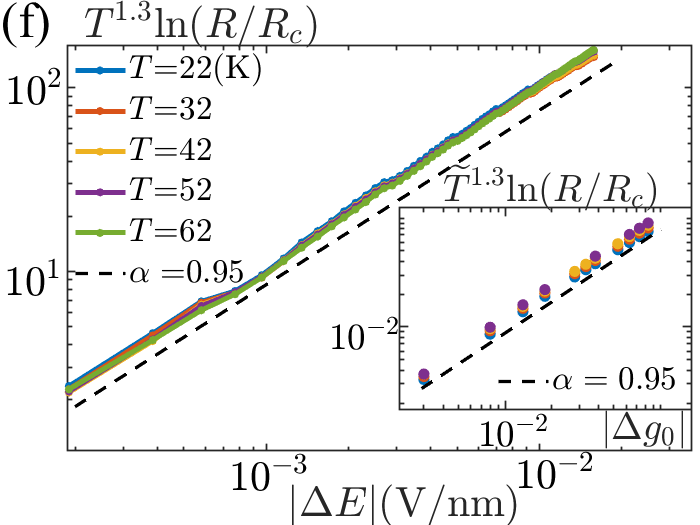}
\caption{Evolution of the resistance for a model (a) without and (b)-(c) with critical resistors. In scenario (b) the insulators have an infinite resistance, while in (c) the insulators are thermally activated. At $T=0$ (violet), the resistance grows continuously upon approaching the bond-percolation threshold without any notion of a jump. The resistance curves in (b) and (c) cross at multiple points (black arrows). (d) The percolation threshold is determined by the total fraction, $f_{\tn{critical}}+f_{\tn{metal}}$, where $f_{\tn{critical}}$ is determined by the typical width of the quantum critical region. The point $g=g_c$ is temperature-independent, as it is defined as the critical point associated with the clean system with a well-defined ``single'' crossing point. The point $g=g_{0}$ associated with the percolation threshold varies as a function of temperature, which is reflected in the horizontal-shift of the resistance curves at different temperatures. (e) The calculated temperature-dependence of resistance as a function of $g_0$. The black dashed line denotes the experimentally observed power-law $(\sim T^{-1.2})$ \cite{li_continuous_2021} and the critical $g_{0,c} \sim -0.35$ (red; see text). Inset: Scaling collapse for $R(T)/R_c(T)$. The color-bar represents a range $-0.1\leq g_0 -g_{0,c}\leq 0.04$. (f) A scaling collapse of the experimental data \cite{li_continuous_2021} and numerical RRN results (inset), as described by the modified functional form in (Eq.~\ref{eq:arrhenius}). The rescaled temperature in the inset is defined as, $\widetilde{T}=k_BT/\Delta_0$, which lies in a range $0.1\leq \widetilde{T}\leq 0.17$.}
\label{fig3}
\end{figure}

\end{widetext}

{\it Results.-} We begin by presenting the evolution of the resistance across the MIT, including the finite-temperature effects (measured in units, $\widetilde{T}=k_BT/\Delta_0$, where $\Delta_0$ is a fixed gap scale deep in the insulator). We consider a semi-infinite strip-like geometry with dimensions, $L_x\times L_y\sim 10^5\times100$. The results in Fig.~\ref{fig3} are obtained for $\rho_c = 5\rho_Q$ and a constant $\rho_m =0.01\rho_Q$; we have obtained similar results for a distribution of metallic resistance values centered around a mean $\rho_m$. Importantly, the key differences arise depending on whether the critical resistors are included in the computation. We note that our coarse-grained percolation model ignores a number of microscopic ingredients, e.g. the explicit temperature dependence of $\rho_m$. However, as our analysis will show, even this simplified treatment captures the essential phenomenology near the MIT, where the dominant effect of increasing temperature is to enter the critical regime (Fig.~\ref{fig2}(a)) with a large $O(\rho_Q)$ resistance. 

First, we consider the simplest scenario without any critical resistors, where the temperature dependence enters via thermal activation. In Fig.~\ref{fig3}(a), we present the evolution of resistance as a function of $g_0$ at different temperatures. It is clear that at any fixed value of $g_0$, the resistance should decrease as a function of increasing temperature due to thermal activation. Importantly, in this scenario there are no crossings, unlike the experimental data. Turning next to the setup including critical resistors, we show results for resistance in Fig.~\ref{fig3}(b) and (c) in the absence and presence of thermally activated insulators, respectively. The traces corresponding to $T=0$ in both curves (violet), where $f_{\tn{critical}}=0$, are identical; the resistance diverges continuously upon approaching the percolation threshold \cite{stauffer}. Within this setup, the notion of a universal resistivity jump at the MIT even at asymptotically low temperatures is ill-defined. Notably, with increasing temperature, the resistance curves exhibit multiple crossing points (black arrows) just as the experimental results in Fig.~\ref{fig1}(c), which is tied crucially to the presence of the critical resistors with a temperature dependent, $f_{\tn{critical}}$ \cite{si}.

To understand the origin of multiple crossings, recall that $p_c$ is determined by the {\it total} fraction of resistors, $(f_{\rm{critical}}+f_{\rm{metal}})$. At $T=0$, $f_{\tn{metal}}$ is determined by the blue region in the top panel of Fig.~\ref{fig3}(d). At finite $T$, $f_{\rm{critical}}\sim T^{1/\nu z}$, the width of the QC fan associated with the underlying clean fixed-point. Thus, at finite $T$, $p_c$ is determined by the total area of the blue and red regions in the bottom panel of Fig.~\ref{fig3}(d). As a function of increasing temperature, as the area of the red region increases, the position of the critical point $g_0$ also changes. This leads to the multiple crossing points.

Turning now to the temperature dependence of the resistance in the scenario including the thermally activated insulators, we show our results as a function of varying $g_0$ in Fig.~\ref{fig3}(e). We identify the critical point as $g_{0,c}\sim -0.347$ (red curve) based on the following two measures. First, one can distinguish between a metallic vs. insulating-like response at low-temperatures depending on the sign of $dR/dT$, with $g_{0,c}$ as an inflection point. Secondly, we can observe scaling collapse for the curves near $g_{0,c}$ onto metallic and insulating branches (inset of Fig.~\ref{fig3}(e)), by considering the normalized resistance, $R(g_0,T)/R_c(T)$, as a function of $T/T_0$. Here, $R_c (T)$ is the resistance at the critical field and $k_BT_0 =\Delta_0 |g_0 -g_{0,c}|^{\nu z}$ represents the charge gap. We observe that the resistance in the nominally metallic low-temperature regime increases and crosses over into an insulating-like response at high-temperatures, in qualitative agreement with the experimental results (Fig.~\ref{fig1}d). This is a consequence of the inclusion of critical resistors along the current path at high temperatures. Remarkably, the asymptotic high-temperature power-law behavior in this scenario is also similar to the experiments, $R_c(T)\sim 1/T^\alpha$, with $\alpha\approx1.2$. It is worth noting that in the scenario without thermal activation, as in Fig.~\ref{fig3}(b), the asymptotic high-temperature behavior of $R_c(T)\sim 1/T^{\alpha'}$ with $\alpha'\approx2.3$ is markedly different \cite{si}. Thus, the scenario where temperature affects both the fraction of critical resistors and activated insulating resistors, appears to show better quantitative agreement with the experiments. To estimate the approximate disorder strength, we fit our $T=0$ results to the experimental result at the lowest temperature, leading to an estimated bound $\sigma \lesssim 5 ~{\rm mV/nm}$ \cite{si}. 

Finally, let us address the behavior of transport at finite temperature near $g_{0,c}$, but on the insulating side of the critical point. If one is not situated in the immediate vicinity of the critical point, and over a small range of intermediate temperatures, the resistance is well described by the usual activated behavior. After all, this is how the charge-gap is extracted in the experiment. However, for a given realization of disorder, our analysis has already shown that the current path involves insulating links (with different local gaps) that are thermally activated, as well as temperature-dependent critical resistors. The latter effect plays an important role in giving rise to the power-law envelope, $R_c(T)$, at higher temperatures. Therefore, it is natural to consider the possibility of describing the full temperature dependence of the resistance, which interpolates between the power-law temperature dependence at $g=g_{0,c}$ and an activated form deep inside the insulator, using a single function,
\begin{equation}
\label{eq:arrhenius}
    R(g_0,T) = R_c (T) \exp \left(\frac{A|g_0 -g_{0,c}|^{\alpha}}{T^{\gamma}}\right),
\end{equation}
where $A$ is a $T$-independent constant and $\alpha, \gamma$ are new exponents. The rationale for this modified form is that it arises due to the flow of the current that averages over insulating links with a distribution of different gap-sizes at low temperatures, and is controlled by the critical resistors at high temperatures. We note that $\alpha$ and $\gamma$ need {\it not} be equal to the expectations from the clean theory, with $\alpha=\nu z$ and $\gamma=1$, respectively. In Fig.~\ref{fig3}(f), by plotting $[T^\gamma \ln(R/R_c)]$ as a function of $|g_0 -g_{0,c}|$, we demonstrate that the above form describes the experimental data \cite{li_continuous_2021} as well as our numerical results (inset) on the insulating side over an extended range of temperatures. Remarkably, we find that choosing $\gamma\approx 1.3$ leads to a collapse of all the curves over a wide range of temperature with $\alpha \approx 0.95$ for both the experimental and numerical data.

{\it Outlook.-} Our work illustrates that a minimal model of long-wavelength disorder induced smearing of a continuous bandwidth-tuned MIT at half filling can account for a number of mysteries associated with recent transport experiments \cite{li_continuous_2021}. However, a number of questions remain. For instance, the experiments report evolution of the spin-susceptibility across the same transition and as a function of temperature. Since the low-temperature susceptibility does not change significantly across the transition, even within a percolation-based picture the susceptibility for the insulating puddles should be comparable to the metallic puddles. If this is not the case, the decreasing fraction of metallic puddles below $p_c$ would lead to a decreasing susceptibility. Similarly, the temperature-dependent evolution of the susceptibility, and the associated Pomeranchuk effect, can arise due to the combined effect of temperature dependence from metallic/insulating puddles and the critical puddles. We leave a detailed theoretical modeling of these observations for future work. Future measurements of the local compressibility and NMR will lead to further insights into the meso-scale inhomogeneities near the MIT. These experiments likely also hold the key for explaining the origin of the difference in the behavior of electrical transport near a similar continuous MIT, but in a moir\'e TMD homobilayer \cite{ghiotto_quantum_2021}.

\acknowledgements

We thank Kin-Fai Mak and Jie Shan for stimulating discussions, and Tingxin Li for sharing with us the experimental data in Fig.~\ref{fig1}. 
SK and DC are supported by Grant No. 2020213 from the United States Israel Binational Science Foundation (BSF), Jerusalem, Israel. TS was supported by US Department of Energy grant DE- SC0008739, and partially through a Simons Investigator Award from the Simons Foundation. This work was also partly supported by the Simons Collaboration on Ultra-Quantum Matter, which is a grant from the Simons Foundation (651446, TS).

\bibliographystyle{apsrev4-1_custom}
\bibliography{refs}

\clearpage

\begin{widetext}

%%%%%%%%%%
%\appendix 
\setcounter{page}{1} 
\renewcommand{\figurename}{Supplemental Figure}
\renewcommand{\theequation}{\thesection.\arabic{equation}}
%%%%%%%%%%%%%%%%

\begin{center}
    {\bf Supplementary material for ``Continuous Mott transition in moir\'e semiconductors: \\role of long-wavelength inhomogeneities"}
\end{center}

%\section{Supplementary Information for ``Continuous Mott transition in moir\'e semiconductors: \\role of long-wavelength inhomogeneities"}

\section{Finite-size scaling of conductance}

\begin{figure}[h]
\centering
\includegraphics[width=0.49\linewidth]{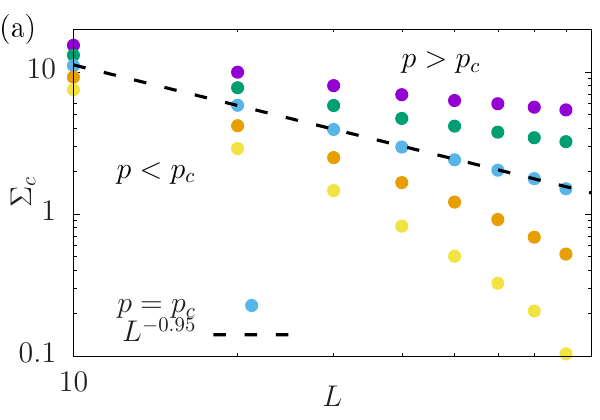}
\includegraphics[width=0.49\linewidth]{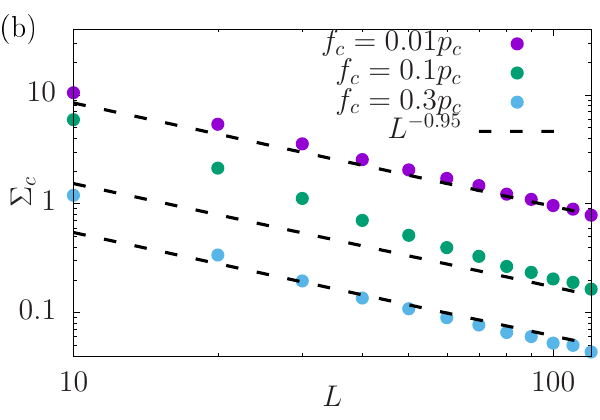}
\caption{Finite-size scaling of conductance in units of \(\rho_Q^{-1}\). (a) The conductance at the percolation threshold $\Sigma_c\sim L^{-0.95}$ for $f_{\tn{critical}}=0$  \cite{stauffer}. (b) For $f_{\tn{critical}}>0$, $\Sigma_c$ displays a crossover to the same power-law regime at increasing $L$.}
\label{fig-appdx-sigc}
\end{figure}

In this section, we discuss the finite-size scaling of the conductance at the bond-percolation threshold. We have also used a variety of numerical values for the ratios of the resistances, which do not affect our conclusions qualitatively. In standard bond-percolation, the conductance behaves as $\Sigma_c \sim L^{-\mu/\nu}$. Here, $\nu=4/3$ is a critical exponent for bond-percolation and $\mu$ is a new exponent; our results in Fig.~\ref{fig-appdx-sigc}(a) are consistent with $\mu/\nu\sim0.95$ \cite{stauffer}. With increasing $f_\tn{critical}$, the system size dependence gets modified at intermediate system-sizes, as in Fig.~\ref{fig-appdx-sigc}(b). However, this is merely a crossover phenomenon, since the power-law behavior associated with usual bond-percolation is restored at large $L$. Recall that $p$ here is the total fraction of metallic and critical resistors, respectively, with $p_c \sim 0.347$ on the triangular lattice. The result for varying $f_{\tn{critical}}$ in Fig.~\ref{fig-appdx-sigc}(b) are obtained for a fixed $\rho_c =100\rho_m$. In the limit of a large system-size, $\Sigma_{c}^{-1}$ is approximately set by the average resistance of the percolating cluster, i.e. $\Sigma_{c}^{-1} \sim f_{\tn{critical}}\rho_c +p_c \rho_m$.

\section{Random resistor network in the absence of thermally activated insulators}

\begin{figure}[h]
\centering
\includegraphics[width=0.5\linewidth]{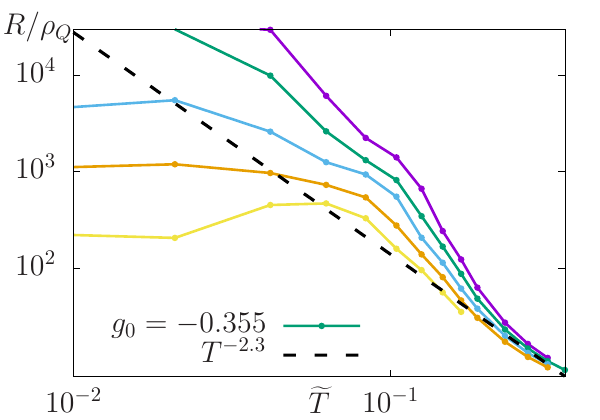}
\caption{Temperature-dependence of resistance in the absence of thermally activated insulators. The black dashed line denotes the power-law \(\sim T^{-2.3}\).}
\label{fig-appdx-noact}
\end{figure}

In this section, we discuss the origin of the asymptotic power-law behavior of the resistance at high temperature in the absence of thermally activated insulating links. In Fig.~\ref{fig-appdx-noact}, we show the temperature dependence of the resistance across the MIT on a log-log plot, which displays a power-law $R_c(T) \sim 1/T^{\alpha'}$ with $\alpha'\approx 2.3$. The origin of this behavior is due to the explicit dependence of $f_{\tn{critical}}$ on $T$, combined with the detailed behavior of the conductance near the percolation threshold, $\Sigma\sim (p-p_c)^{\mu}$. To capture this asymptotic power-law behavior at high temperature in the critical regime, we approximate
\beq
\Sigma \sim (p(g_0,T)-p_c)^{\mu} \sim T^{\mu/(\nu z)},
\eeq
where $\nu, z$ are the critical exponents that we have explicitly assumed, associated with the bandwidth-tuned MIT in the clean theory. With $\mu=4/3$ and $\nu z\approx 0.67$, the above combination leads to a value that is close to $\alpha'$. Importantly, this result is markedly different from the experiment $R_c(T)\sim T^{-1.2}$, which agrees better with the RRN results once the effects of thermally activated insulators are included. In the latter description, the network of insulating links with a distribution of gap-sizes near the backbone, contribute to the conductivity.

\section{Effective medium theory}

\begin{figure}[h]
\centering
\includegraphics[width=0.49\linewidth]{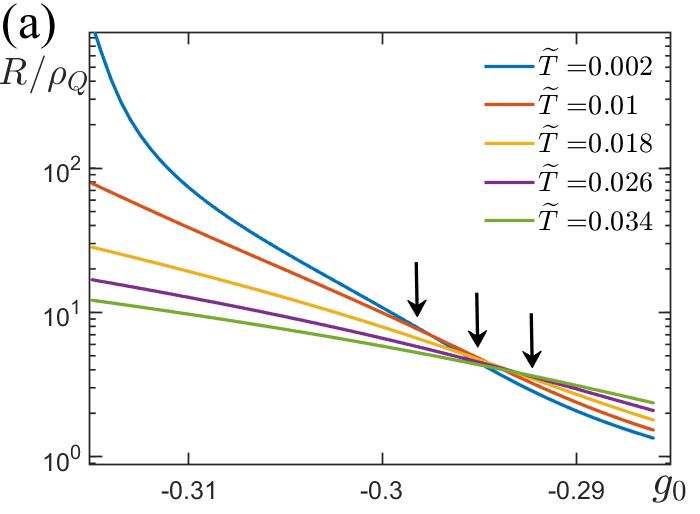}
\includegraphics[width=0.49\linewidth]{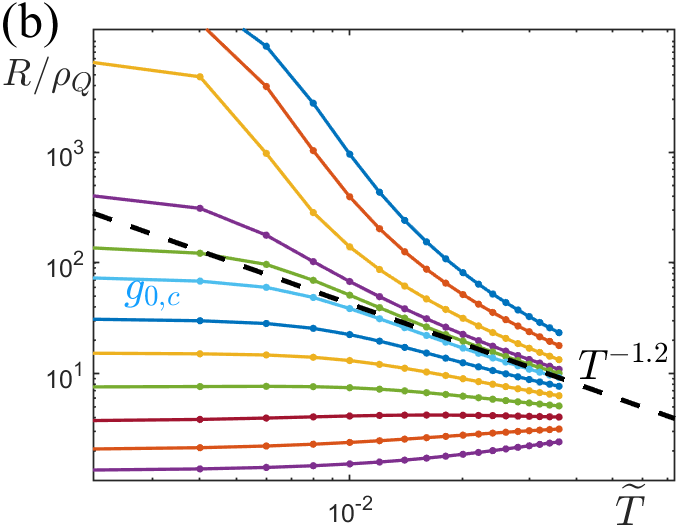}
\includegraphics[width=0.49\linewidth]{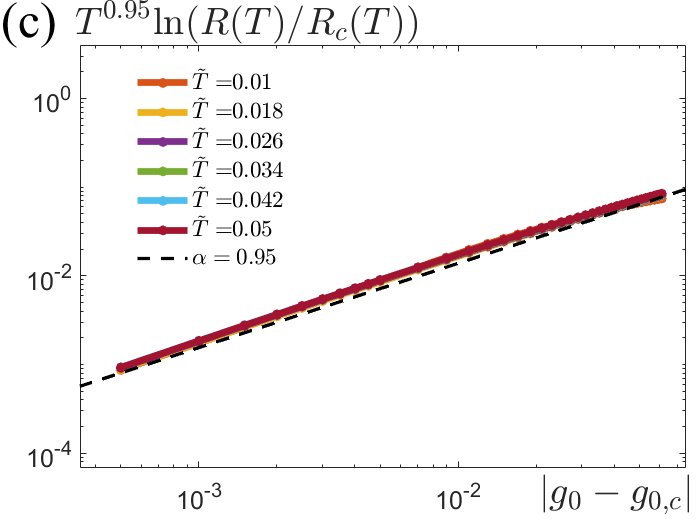}
\includegraphics[width=0.49\linewidth]{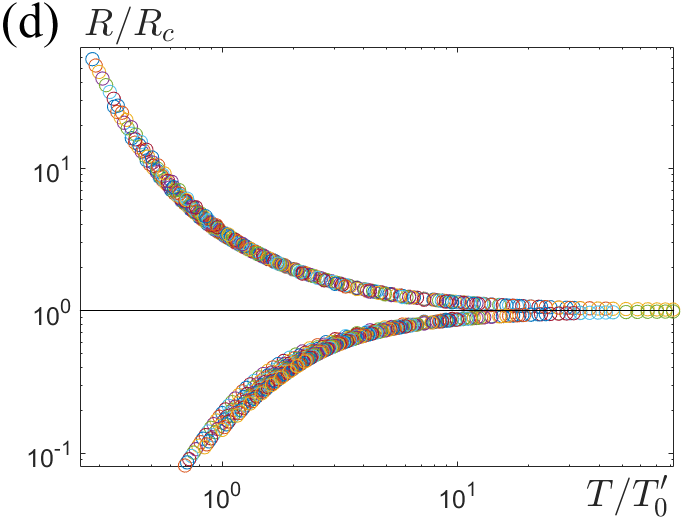}
\caption{The numerical results for EMT calculations. (a) Field-dependence of resistance. Black arrows indicate multiple crossing points. (b) Temperature-dependence of resistance. \(g_{0,c}\sim -0.31\) is the critical field and the black dashed line marks the power-law observed in the experiment. (c) Modified gap activation form for the resistance with fitting parameters \(\alpha=0.95, \gamma=0.95\) (defined in the main text Eq.~\ref{eq:arrhenius}). (d) Scaling collapse for \(R(T)/R_c(T)\) as a function of a normalized temperature \(T/T_{0}'\). \(T_{0}' \sim |g_0 -g_{0,c}|^{\alpha}\) indicates the energy scale defined in the modified gap activation.}%(d) Scaling collapse near \(g_{0,c}\) when the critical exponents of the clean theory \((0.67)\) are used.}
\label{fig-appdx-emt}
\end{figure}
%\end{widetext}

In this section, we present results for the resistivity obtained using effective medium theory (EMT), which is qualitatively consistent with the RRN results obtained earlier. The EMT is an alternative approach for computing the conductance of a disordered system. The basic setup is similar, with the system described in terms of puddles with a Gaussian distribution centered around a mean value of the tuning parameter, $g_0$. The effective conductance, $\Sigma_e$, within EMT satisfies a self-consistent equation \cite{Redner_2012} %\cite{torquato2002random}
\beq
\label{eq:emt_sca}
    \sum_{g}P_{\sigma}(g,g_0) \frac{\Sigma(g) -\Sigma_e}{\Sigma(g) +x\Sigma_e}=0,
\eeq
where $g$ sums over the entire distribution of puddles, $P_{\sigma}(g,g_0)$ is the disorder distribution as earlier, and $\Sigma(g)$ is the conductance of the puddle. $x=2$ is a geometry-dependent constant tied to the underlying triangular lattice. In Fig.~\ref{fig-appdx-emt}, we show our EMT results for the same parameter values that were chosen for the RRN computations, $\rho_c =5\rho_Q, \rho_m =0.01\rho_Q$. Unsurprisingly, the resistance evolution continues to display multiple crossing points in Fig.~\ref{fig-appdx-emt}(a). 

In Fig.~\ref{fig-appdx-emt}(b), we present the results as a function of temperature. As with RRN, one can identify a critical field $(g_{0,c}\sim-0.31)$ by a scaling analysis, and also describe the complete temperature dependence on the insulating side in terms of Eq.~\ref{eq:arrhenius} in the main text (Fig.~\ref{fig-appdx-emt}(c)). While the exponent $\alpha\approx0.95$ is still the same as RRN and the actual experiment, $\gamma$ obtained from EMT is slightly different and given by $0.95$. Using the temperature dependent $R_c(T)$, we can obtain mirror symmetric branches from the scaling collapse plot as shown in Fig.~\ref{fig-appdx-emt}(d).

\section{Estimating the strength of long-wavelength disorder}

\begin{figure}[h]
\centering
\includegraphics[width=0.5\linewidth]{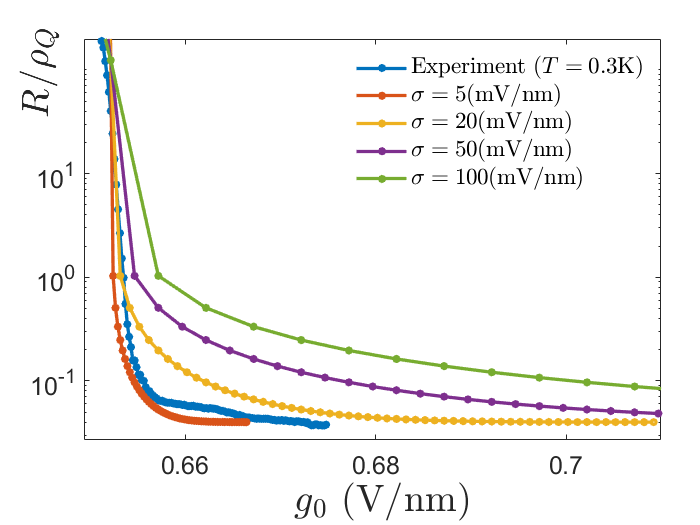}
\caption{Estimation of the disorder strength of the experiments. We use experimental data obtained at the lowest temperature (\(T\sim 0.3\tn{K}\)), at which critical resistors should be absent. We choose the resistance of a metallic resistor as \(\rho_m \sim0.04\rho_Q\), which is a convergent value of the experimental resistance deep in the metallic side. The disorder strength is roughly estimated as \(\sigma \lesssim 5\tn{(mV/nm)}\).}
\label{fig-appdx-disorder}
\end{figure}

The disorder strength in the experiments can be estimated from the resistance evolution at the lowest temperature. Assuming that the resistance evolution at the lowest temperature \((T\sim0.3\tn{K})\) is well described by a two-phase bond-percolation without critical resistors, we obtain \(\rho_m =0.04\rho_Q\) from the experiments deep in the metallic side \((g_0 \sim 0.675\tn{(V/nm)})\). Fig.~\ref{fig-appdx-disorder} shows the experimental data and EMT results in the \(T\rightarrow 0\) limit at various disorder strengths. We can see that a disorder strength \(\sigma > 5(\tn{mV/nm})\) leads to a significant discrepancy between the experimental and the numerical results. Thus, on this crude comparison, we estimate that the disorder strength should be \(\sigma \lesssim 5\tn{(mV/nm)}\).   

\section{Recovery of the usual activated behavior}

\begin{figure}[h]
\centering
\includegraphics[width=0.49\linewidth]{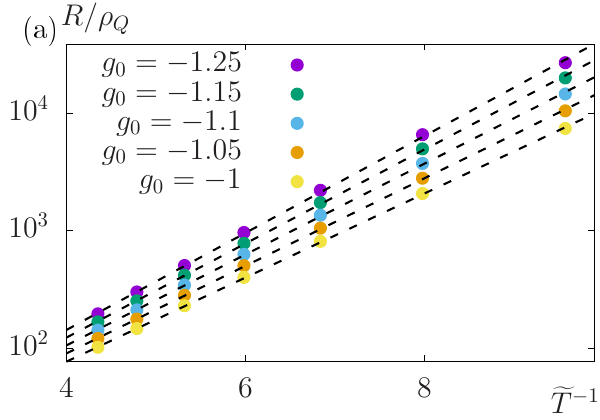}
\includegraphics[width=0.49\linewidth]{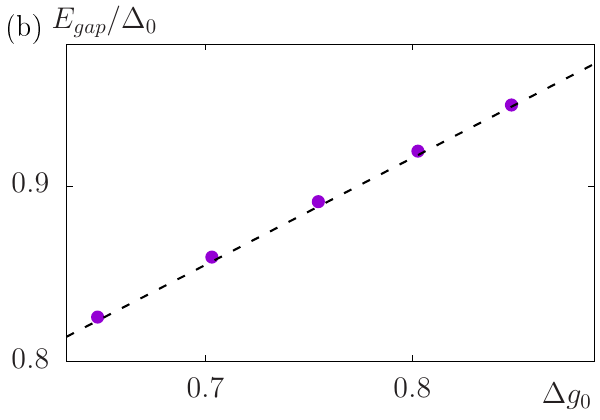}
\includegraphics[width=0.49\linewidth]{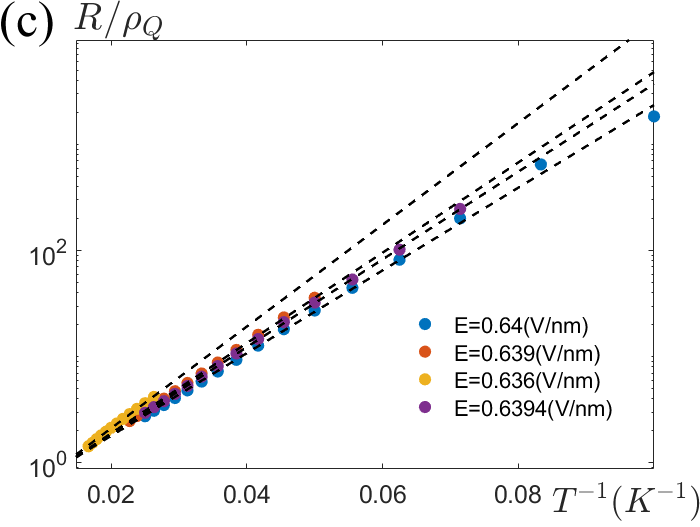}
\includegraphics[width=0.49\linewidth]{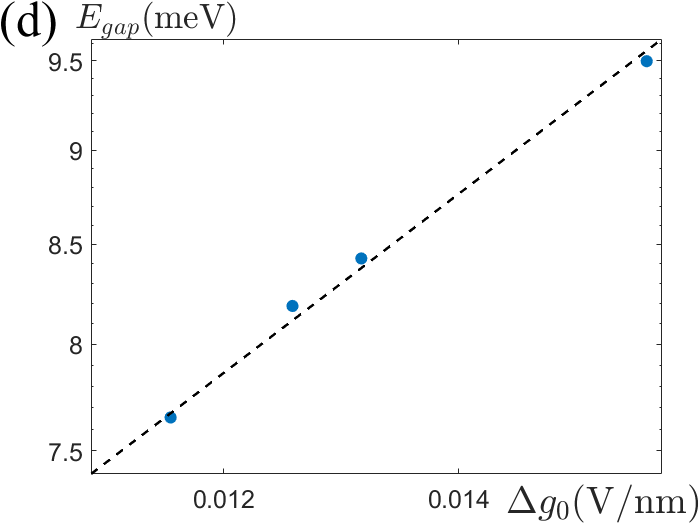}
\caption{Recovery of the usual thermal activation form over an intermediate regime of temperatures. (a) Extraction of activation gap from the RRN results. The black dashed lines show \(\tn{ln}R \sim T^{-1}\) behaviors, where the slope denotes the activated gap. (b) The activated gaps as a function of \(\Delta g_0\) on a log-log plot. The slope of the black dashed line stands for \(\alpha_\tn{disordered}=0.54\pm 0.01\). %which is very close to \(\alpha_\tn{clean}=0.67\). (c) and (d) show the extraction of activation gap from the RRN results, which give \(\alpha_\tn{disordered}=0.73\pm0.03\). 
(c) and (d) show the gap extraction from the experimental results, from which we find \(\alpha_\tn{disordered}=0.70\pm0.03\).}
\label{fig-appdx-usual-gap}
\end{figure}

Near the MIT, the thermal activation form needs to be modified to comply with the temperature dependence of \(R_c(T)\) as shown in the main text. However, the usual activation form should be valid away from the immediate vicinity of the MIT, over an intermediate regime of temperatures. The upper bound of the temperature regime can be set by the crossover temperature \(T^* \sim |\Delta g_0|^\alpha\), below which the resistance should be reasonably well described by an activation form . Here \(\Delta g_0\) is measured from the usual percolation threshold. Additionally, the usual activation behavior cannot persist down to the lowest temperature because, at low temperatures, the resistance curves in the insulating side show a saturation behavior due to the inclusion of metallic puddles, as shown in Fig.~\ref{fig3}(e) and Fig.~\ref{fig-appdx-emt}(b). In Fig.~\ref{fig-appdx-usual-gap}(a), we plot the RRN resistance values as a function of \(\widetilde{T}^{-1}=\frac{\Delta_0}{k_B T}\) over an intermediate regime of temperatures described above. Here, the slope of the black dashed lines shows an extracted gap. We present the extracted gap values as a function of \(\Delta g_0\) on a log-log plot in Fig.~\ref{fig-appdx-usual-gap}(b) to obtain the slope \(\alpha_\tn{disordered}=0.54\pm0.01\), We also plot extracted gap values for the experimental results in Fig.~\ref{fig-appdx-usual-gap}(c), and obtain the slope \(\alpha_\tn{disordered}=0.70\pm0.03\) in Fig.~\ref{fig-appdx-usual-gap}(d).

\end{widetext}
\end{document}